\documentclass[]{pasj01}
\draft

\begin{document}
\SetRunningHead{Author(s) in page-head}{Running Head}
\Received{2015 April 14}
\Accepted{2015 December 28}{
\Published{2016 March 2}

\title{A Soft X-Ray Lag Detected in Centaurus A}

\author{Yutaro Tachibana \altaffilmark{1, *},
		Taiki Kawamuro \altaffilmark{2},
		Yoshihiro Ueda \altaffilmark{2},
		Megumi Shidatsu \altaffilmark{3},
		Makoto Arimoto \altaffilmark{1},
		Taketoshi Yoshii \altaffilmark{1},
		Yoichi Yatsu \altaffilmark{1},
		Yoshihiko Saito \altaffilmark{1},
		Sean Pike \altaffilmark{4},
		and
		Nobuyuki Kawai \altaffilmark{1}
		}

\altaffiltext{1}{Department of Physics, Tokyo Institute of Technology, 2-12-1, Ohokayama, Tokyo, Japan}
\altaffiltext{2}{Department of Astronomy, Kyoto University, Oiwake-cho, Sakyo-ku, Kyoto 606-8502, Japan}
\altaffiltext{3}{MAXI team, Institute of Physical and Chemical Research, Wako, Saitama 351-0198, Japan}
\altaffiltext{4}{Depertment of Physics, Brown University, Providence, Rhode Island 02912, USA}
 
\email{tachibana@hp.phys.titech.ac.jp}

%


\maketitle


\begin{abstract}
 We performed time lag analysis on the X-ray light curves of Centaurus A (Cen A) 
obtained by the Gas Slit Camera (GSC) aboard the Monitor of All-sky X-ray Image ({\it MAXI})
in three energy bands (2--4 keV, 4--10 keV, and 10--20 keV). 
We discovered a soft X-ray lag relative to higher energies (soft lag) on a timescale of days 
in a flaring episode  
by employing the discrete correlation function (DCF) 
and the z-transformed discrete correlation function (ZDCF) method. 
In the episode, 
a peak and a centroid in the DCF and the ZDCF was observed at a soft lag of $\sim 5$ days in 2--4 keV versus 4--10 keV and 
in 4--10 keV versus 10--20 keV, and $\sim 10$ days in 2--4 keV versus 10--20 keV. 
We found it difficult to explain the observed X-ray variation 
by a single energy injection 
with the one-zone synchrotron self-Compton (SSC) model,
in which the soft lags in these three energy bands reflect the different cooling times of the relativistic electrons, 
by assuming the magnetic field and minimum Lorentz factor 
estimated from a broad-band SED. 
Alternatively, if the phenomenon is interpreted as cooling of Comptonizing electrons in a corona covering the accretion disk, 
the temperature of the corona producing the variable X-rays should be $\sim$10 keV 
for reconciliation with the soft lag in the energy range of 2--20 keV.
\end{abstract}


\section{Introduction}
Centaurus A (Cen A, NGC 5128), at a distance of 3.8 Mpc (Harris et al. 2010), is the nearest active galactic nuclei (AGN) 
from the Milky Way galaxy and is known as the brightest Seyfert galaxy in the X-ray sky 
($2.87 \times 10^{-10}\ \rm{erg\ cm}^{-2}\ \rm{s}^{-1}$ in the 4--10 keV band; Hiroi et al. 2011, and 
$7.48 \times 10^{-10}\ \rm{erg\ cm}^{-2}\ \rm{s}^{-1}$ in the 14--195 keV band; Tueller et al. 2008). 
In addition, this object has often been considered as a jet-dominated AGN seen from the side called ``misdirected BL Lac object'' 
(Morganti et al. 1992; Chiaberge et al. 2001) with its jet inclined to the viewing angle at $\gtrsim 50^{\circ}$ (Tingay et al. 1998). 
Owing to its brightness and proximity, this object has been an ideal target to study the geometry and the emission mechanism 
around a supermassive black hole. 
However, the origin of the X-ray radiation from Cen A has not been easily identified 
from its energy spectrum. 
The observed X-ray spectrum has no significantly different features from typical Seyfert galaxies, which radiate X-rays mainly from 
the inner disk region with Comptonization in hot plasmas located above or around the accretion disk 
({\it e.g.} Rothschild et al. 1999; Beckmann et al. 2011; Fukazawa et al. 2011). 
On the other hand, the high energy tail of the X-ray spectrum appears to be smoothly connected to the MeV emission 
which is believed to be generated from a jet, and the spectral energy distribution (SED) in the energy range between
radio and GeV $\gamma$-ray can be modeled successfully by the one-zone SSC model or other varieties of jet emission models 
({\it e.g.} Abdo et al. 2010; Roustazadeh \& B$\ddot{\rm{o}}$ttcher 2011; Petropoulou et al. 2013), 
though a combination of a thermal inverse Compton component and a non-thermal jet emission 
also describe the SED equally well (Abdo et al. 2010). 

When the spectral analysis does not provide a unique emission mechanism 
or physical parameters of AGNs, 
time variability analysis, especially time lag analysis, may be useful to 
obtain additional information to resolve the issue. 
Kataoka et al. (2000) detected soft lags 
({\it i.e.} the variation in a lower energy band lags respect to that of a higher energy band)
on a timescale of $\sim 10^3$--$10^4$ sec in an X-ray flare from the TeV blazar PKS 2155-304 
and estimated the magnetic field strength at the X-ray emitting region in the jet with the hypothesis 
that the delays are caused by different synchrotron lifetimes of the relativistic electrons. 
Similar discussion was presented by Tanihata et al. (2001) for other blazar objects: Mrk 421 and Mrk 501. 
Sriram et al. (2009), conversely, found hard lags (vice versa for soft lag) on a timescale of $\sim 10^2$--$10^3$ sec 
in the Seyfert 1 galaxy NGC 4593 using the {\it XMM-Newton}/EPIC. 
The authors interpreted the hard lags as a timescale of thermal Compton scattering in a compact central corona 
and derived its size as about five Schwarzschild radii ($R_{\rm s}$) for the black hole mass $7 \times 10^6 M_{\odot}$. 
Recently, De Marco et al. (2013), based on the study of 32 radio quiet AGNs, discovered the relation 
between black hole masses ($M_{\rm{BH}}$) and soft X-ray lags ($\tau \approx 10$--600 sec) 
in high-frequency variation ($\nu \approx 0.07$--$4 \times 10^{-3}$ Hz); 
$\rm{\log}|\tau| = 1.98[\pm 0.08] + 0.59[\pm 0.11] \rm{\log}(M_{\rm{BH}}/10^7M_{\odot})$. 

For Cen A, time lag analysis in the X-ray band has scarcely been applied as yet, 
presumably because of remarkably stable spectral (or flux) behavior 
as reported by Rothschild et al. (2011) based on 12.5-year observation by the {\it Rossi X-ray Timing Explorer} 
combined with pointed observation with inadequate time coverage or continuous monitoring with insufficient sensitivity 
in the X-ray band. 
{\it Monitor of All-sky X-ray Image} ({\it MAXI}; Matsuoka et al. 2009), launched in August 2009, 
has a high sky coverage of ~95\% 
and a X-ray sensitivity to detect above 15 mCrab in the 4--10 keV band with a cadence of $\sim$90 min (Sugizaki et al. 2011), 
and thus allowing us to investigate a long-term activity for AGNs with a sufficient time resolution. 

 In this paper, we study X-ray variation properties in Centaurus A.
The lags in long-look flux variations are investigated using light curve data sets obtained by the {\it MAXI}/GSC in MJD = 55058--56800 in \S4.1.
Then we focus on a flare-like enhancement event observed during MJD = 55075--55160 in \S4.2. 
For the lag analyses, we employed the continuously evaluated discrete correlation function (CEDCF)
based on the discrete correlation function (DCF; Edelson \& Krolik, 1988), 
and the z-transformed discrete correlation function (ZDCF; Alexander, 1997). 
Finally, we discuss the interpretations of this phenomenon and the origin of X-ray emission of Cen A in \S5.


\section{Observations and Light Curves}
In August 2009, {\it MAXI} was launched and attached 
to the Japanese experiment module ``Kibo'' on the {\it International Space Station} ({\it ISS}).  
Since then, the gas slit camera (GSC; Mihara et al. 2011) and solid-state slit camera (SSC; Tomida et al. 2011) aboard {\it MAXI} 
has been producing an all sky X-ray image at every orbital period of {\it ISS} (92 minutes). 
The SSC employs X-ray CCD arrays and covers the energy range 
from 0.5 keV to 12 keV, while, the GSC employs gas proportional counters and covers 
the energy range from 2 keV to 20 keV. 
Figure \ref{fig:lc_all} shows the {\it MAXI}/GSC light curve during MJD=55058--56800 in the three energy bands 
obtained by the {\it MAXI} on-demand analysis (ver 2.0) implemented at the {\it MAXI} public web site\footnote[1]{http://maxi.riken.jp/top/}, 
where calibration  described in Nakahira et al. (2012) is applied. 
These light curve are binned by five days. 
The flux histories of Cen A reveal variations of a factor of  $\sim$3--5 on a timescale of the order of several weeks or months.
In this analysis, we used only the bin meeting the condition that 
the data whose error size $e_i$ is smaller than $\bar{e_i} + 3\sigma  _{e_i}$, 
where the parameter $\bar{e_i}$ is the mean of $e_i$, and $\sigma_{e_i}$ is the standard deviation of $e_i$. 
The purpose of this treatment is removing the low quality data 
resulting from a relatively insufficient number of scans in the five day interval.
5--7\% of the data points were removed by the treatment described above. 
It should be noted that the sensitivity of the {\it MAXI}/GSC 
has deteriorated over time due to damages in the detectors and changes of the operation modes. 
For more information, see Mihara et al. (2014).

\section{Method of Time Lag Analysis}

To evaluate time lags in X-ray variation between two different energy bands, we employed two techniques based on the discrete correlation function, 
called the continuously evaluated discrete correlation function and the z-transformed discrete correlation function. 

\subsection{Discrete correlation function}
The DCF was introduced by Edelson and Krolik (1988). 
To analyze the irregularly sampled astronomical data, 
DCF has been used widely to uncover correlations or lags between flux variations in two different energy bands, especially for AGNs.
This function provides the correlation coefficient between two unevenly sampled time-series data as a function of lag ($\tau_{\rm{lag}}$), 
and the value $\tau_{\rm{lag}}$ corresponding to the peak or the centroid of $DCF(\tau_{\rm{lag}})$, 
is presumed to be the value of the lag in the variations in two time series. 
First, we calculate the set of un-binned discrete correlation functions (UDCFs) for all measured pairs ($a_i, b_j$) defined as follows: 
\begin{equation}
UDCF_{ij}  = \frac{a_i - \bar{a}}{ \sigma_a } \times \frac{b_j - \bar{b}}{ \sigma_b } ,
\end{equation}
where $\bar{a}$ and $\bar{b}$ are the means, and $\sigma _a$ and $\sigma _b$ are the standard deviation of data sets $a$ and $b$. 
Each $UDCF_{ij}$ is associated with the pairwise lag $\Delta t_{ij} = t_j - t_i$. 
Next, we average the $M$ pairs for which over $\tau_{\rm{lag}} - \Delta \tau_{\rm{lag}}/2 \le \Delta t_{ij} < \tau_{\rm{lag}} + \Delta \tau_{\rm{lag}}/2$, 
\begin{equation}
DCF(\tau_{\rm{lag}}) = \frac{1}{M} \sum_{ij} UDCF_{ij}.
\end{equation}
One $\sigma$ error of each bin is defined by the following expression:
\begin{equation}
\sigma(\tau_{\rm{lag}}) = \frac{1}{M-1} \left[\sum_{ij} \left( UDCF_{ij} - DCF(\tau_{\rm{lag}}) \right)^2 \right]^{1/2}.
\end{equation}

\subsection{Continuously evaluated discrete correlation function}
The continuously evaluated discrete correlation function (CEDCF) was originally introduced by Goicoechea et al. (1998). 
The CEDCF is an oversampled DCF, meaning that the bin-to-bin interval is smaller than the bin width, and 
thus this method allows us to determine more precise time lags from the position of the peak or centroid than 
via the standard DCF method. 
Details of  advantages of this method are described in Gil-Merino et al. (2002). 
Here, we used the centroid instead of the peak lag, because the centroid is less sensitive to a particular binning used 
and statistical fluctuation. 
In order to estimate the significance and uncertainties of the time lags, 
we built a cross-correlation centroid distribution (CCCD) 
by running Monte-Carlo simulations known as ``flux redistribution/random subset selection" (FR/RSS) 
described in detail by Peterson et al. (1998). 
For every simulation, in order to account for the effects of uncertainties in the measured flux, 
we altered each measurement value to a random Gaussian deviate 
by assuming that flux uncertainties are Gaussian distributed about measured value with uncertainty (FR), 
and in order to account for the effects of sampling in time, 
we select random $N$ points  
from $N$ real data points 
regardless of whether they have been previously selected or not
(RSS; known as ``bootstrap method", and this procedure reduces the number of points in each light curve by a factor of $\sim 1/e$). 
This method yields fairly conservative uncertainties in the obtained lags, 
{\it i.e.,} the real uncertainty may be somewhat smaller (Peterson et al. 1998). 
Here, we calculated the centroid of the CEDCF ($\tau_{\rm{cent}}^{\rm{CEDCF}}$) as 
$(\Sigma_i \tau_i CEDCF_i) / (\Sigma_i CEDCF_i )$
using the points with correlation coefficients larger than the half of the peak ($ CEDCF_i \ge 0.5 \times CEDCF_{peak} $). 
In each simulation, the centroid of a CEDCF is determined and recorded to build the CCCD, 
and then we can obtain $\tau _{\rm{median}}$ and $\pm \Delta \tau_{68}$ directly from the constructed CCCD respectively, 
where $\pm \Delta \tau_{68}$ are defined as that 
15.87\% of the realizations give values below $\tau _{\rm{median}} - \Delta \tau_{68}$ 
and above $\tau _{\rm{median}} + \Delta \tau_{68}$ 
({\it i.e.,} 68.27\% of yielded centroids are contained between 
$\tau _{\rm{median}} - \Delta \tau_{68}$ and  $\tau _{\rm{median}} + \Delta \tau_{68}$). 
Thus, the  values  $\pm \Delta \tau_{68}$ correspond to 1$\sigma$ errors for a normal distribution.

\subsection{Z-transformed discrete correlation function}
Alexander (1997) proposed the z-transformed discrete correlation function (ZDCF) to improve the CCF analysis of sparse and 
unevenly sampled light curves. 
Several biases of the conventional DCF method can be corrected by the ZDCF technique by using Fisher's z-transform and 
equal population binning rather than equal time interval $\Delta \tau_{\rm{lag}}$.
If we have $n$ pairs in a given time lag $\tau_{\rm{lag}}$, the correlation coefficient at this lag $r(\tau_{\rm{lag}})$ is estimated by 
\begin{equation}
r(\tau_{\rm{lag}}) = \frac{1}{n-1}\sum_{ij}\left( \frac{a_i - \bar{a}}{\sigma_a} \times \frac{b_j - \bar{b}}{\sigma_b} \right), 
\end{equation}
where $\bar{a}, \bar{b}$ and $\sigma_a, \sigma_b$ are the means and the standard deviations of data sets $a$ and $b$, respectively. 
The sampling distribution of $r$ is highly skewed ({\it i.e}, the distribution is far from normal distribution), 
and therefore calculating its sampling error by simple standard variance of $r$ might be very inaccurate. 
Assuming $a$ and $b$ are drawn from the bivariate normal distribution, 
we can transform $r$ into an approximately normally distributed 
random variable by Fisher's z:
\begin{equation}
z = \frac{1}{2} \log \left ( \frac{1+r}{1-r} \right), \hspace{5mm}
\rho = \tanh z, \hspace{5mm}
\zeta = \frac{1}{2} \log \left( \frac{1+\rho}{1-\rho} \right).
\end{equation}
The mean and the variance of $z$ are approximately equal to 
\begin{equation}
\bar{z} = \zeta + \frac{\rho}{2(n-1)} \times
\left[ 
1+ \frac{5+\rho^2}{4(n-1)} + \frac{11+2\rho^2+3\rho^4}{8(n-1)^2} + \cdots
\right], 
\end{equation}
\begin{equation}
s_z^2 = \frac{1}{n-1} 
\left[ 
1+ \frac{4-\rho^2}{2(n-1)} + \frac{22-6\rho^2-3\rho^4}{6(n-1)^2} + \cdots
\right]. 
\end{equation}
These relations yield the interval corresponding to the $\pm $ 1$\sigma$ error interval which is represented as 
\begin{equation}
\delta r _{\pm} = |\tanh (\bar{z} \pm s_z) - \rho|. 
\end{equation}
The binning is performed with a minimum lag width of $\epsilon$ and a minimum UDCF number of $n_{\rm{min}}$ for each bin, 
where $n_{\rm{min}}$ is at least 11 points for a meaningful statistical interpretation.
Also, the interdependent pairs are discarded in each bin; 
namely a new pair whose $a$ or $b$ points have previously appeared in that bin is not gathered anymore. 
The lag time associated with a bin is defined as the mean value in the bin, and its upper and lower errors ($\Delta \tau_{\pm}$) 
are defined as the intervals that contain 34.1\% of the points in the bin above and below $\bar{\tau}$, respectively. 
It is more robust method than conventional DCF method when the light curves are very sparsely and irregularly sampled. 
The ZDCF time lag is characterized by the centroid $\tau_{\rm{cent}}^{\rm{ZDCF}}$, 
which can be calculated through the same procedure as that of the CEDCF. 

\section{Analysis and Result}

\subsection{Analysis of Complete Time Series}
First, we considered the 4.5 years light curve data sets. 
To search for possible time lags and correlation between each fluctuation in the three energy bands, 
we calculated the CEDCFs($\tau_{\rm{lag}}$) and ZDCFs($\tau_{\rm{lag}}$) for all energy combinations 
(2--4 keV vs. 4--10 keV, 4--10 keV vs. 10--20 keV, and 2--4 keV vs. 10--20 keV) 
for the range of lags between $-200$ days and $+200$ days, 
where the CEDCFs are calculated with the following values: 
5 days as the bin-to-bin interval and 15 days as the bin width $\Delta \tau_{lag}$ (see \S 3.2). 

Results are displayed in Figure \ref{fig:dcf_all}. 
We present the CEDCFs and the error ranges associated with them as the gray shadowed areas, 
and the ZDCFs as black circles with error bars. 
Here the CEDCFs and the ZDCFs show almost the same behavior.
It is to be noted that 
all of the maximum values of the correlation coefficients calculated with the both methods are located at ${\rm Lag} > 0$. 
Again, in this paper, a positive time lag indicates that the variation of soft X-rays is delayed with respect to that of hard X-rays. 
This result, therefore, suggests that a ``soft lag" on timescales of several days exists in flux variations. 
We will examine the significance of the soft lags quantitatively below. 
A similar lag on timescales of a day in the X-ray band has been reported in NGC 4151 by Caballero-Garcia et al. (2012). 
They found the asymmetric shape of DCF
between light curves in the 20--50 keV and 50--100 keV obtained by {\it Swift}/BAT, 
{\it i.e. $DCF(\tau_{\rm lag}>10 {\rm\ days}) $} is larger than $DCF(\tau_{\rm lag}<-10 {\rm\ days}) $,
suggesting that the hard band lead the soft band variations on timescales longer than $\sim$ 10 days. 

To evaluate the statistical uncertainties of the soft lags, we performed Monte-Carlo simulations utilizing the FR/RSS method. 
For constructing CCCDs, we calculated CEDCFs for 1000 light curves produced by Monte-Carlo simulations with a bin size of 10 days. 
In order to minimize the systematic effect of bin size, we performed similar calculations with 15, 20, and 25 day bins, 
and consequently a total of 4000 simulations are performed to obtain a distribution of centroids. 
In the same manner as those for the CEDCF, we also performed simulations for smoothed ZDCF calculated by
using running average of 10, 15, 20, and 25 days, 
where we selected $n_{\rm{min}}=11$ and $\epsilon = 5$ days for each simulated ZDCF in this analysis. 
The distributions of the centroid of CEDCFs and smoothed ZDCFs are presented in Figure \ref{fig:hist_all}. 
Solid lines show the CCCDs derived through the CEDCF method, and dotted lines show 
the CCCDs built with the ZDCF method.
On the basis of these distributions, we calculated the 1 $\sigma$ error ranges of the soft lags. 
The results are consistent between two methods. 
Table \ref{tab:cent_all} shows the magnitudes of the soft lags and associated 1$\sigma$ error estimated. 
The first column gives the energy bands used for the calculation, and the second and fourth give the possible lag 
obtained by the CCCD of a CEDCF $\tau^{\rm{CEDCF}}_{\rm{cent}}$ and a ZDCF $\tau^{\rm{ZDCF}}_{\rm{cent}}$, respectively. 
Based on the CCCD, we also show the probability $P_{<0}$ that the true lag is less than zero in the table.
Here, the realizations of $P_{<0}$ gives 
the probability of the simulated centroids lie in the negative day region,
where soft lag is represented as a positive value. 
In spite of the centroids with highest possibility in 4--10 keV vs. 10--20 keV and 2--4 keV vs. 10--20 keV 
are evaluated to be positive values in both methods, 
their error ranges include the zero, {\it i.e.} no lag, 
and the confidence levels of the soft lags are less than or comparable to 82.8\%. 
The soft lags averaged over the complete time series, therefore, are not highly confident. 

Next, in order to clarify the possible time lags and correlation between each fluctuation in the three energy bands as a function of time, 
we calculated the running CEDCF($t_k$,$\tau_{lag}$) between separated data sets ($a_i, b_j$) contained in the time width 
$t_k - \Delta t / 2 \le t < t_k + \Delta t / 2$ (k = 0,1,2 $\cdot\cdot\cdot$) by employing a rectangular window function. 
The resulting running CEDCF($t_k$, $\tau_{lag}$) is depicted in Figure \ref{fig:box_dcf}. 
Here, we set the parameter $\Delta t $ as 200 days, $\Delta \tau_{lag}$ as 20 days, 
and the distance between centers of bins ($t_{k+1}-t_{k}$) as 5 day. 
The horizontal axis corresponds to the center of the window ($t_k$), and the gray scale denotes the correlation coefficient.
To assist comparison, 
the light curve of the 2--20 keV is also displayed on the upper side.
Until MJD $\sim$ 55800 the CEDCFs have relatively high correlation coefficient 
(represented by dark gray) almost continuously in all CEDCFs.
Here, the peak of the correlation coefficient stay roughly constant in all cases. 
The lag is $\sim$ 0 days for 2--4 keV vs. 4--10 keV, while $\sim$ 10 days for the others. 
These lags are consistent with the results shown in Figure \ref{fig:dcf_all}. 
Especially, the portion between 55075--55150, which contains the highest flux enhancement event,
yielded rahter clear peaks of the CEDCF in all energy combinations. 
On the other hand,  
there are not prominent features in CEDCFs during the period MJD $\sim$ 55800 to MJD $\sim$ 56250,
probably due to insignificant flux fluctuations in each time window of 200 day. 
For MJD $\gtrsim$ 56250, while the coefficients are lager than the previous period as a whole, 
peaks of DCFs are broader and centroids are rather ambiguous. 
Although the changes of the shape and the feature in the CEDCFs might reflect the transition of X-ray emission mechanism in Cen A, 
the episodes in the later epochs are more difficult to study due to reduced sensitivity of the {\it MAXI}/GSC. 
In this parer, we focus on the first flare-like flux enhancement episode, where the correlations are particularly significant. 

\begin{table}
 \label{table:data_type}
 \begin{center}
  \begin{tabular}{lcccc}
   \hline
   Energy band (keV) & $\tau ^{\rm{CEDCF}}_{\rm{cent}}$ (days) & $P^{\rm{CEDCF}}_{<0 \ (\rm{ days})}$
   & $\tau ^{\rm{ZDCF}}_{cent}$ (day) & $P^{\rm{ZDCF}}_{<0 \ (\rm{ days})}$ \\
   \hline \hline
   2--4 vs 4--10  & $0.1 _{-4.5} ^{+4.8}$ & 0.484 & $-0.3 _{-4.6} ^{+4.4}$ & 0.536\\
   4--10 vs 10--20  & $1.3 _{-4.5} ^{+4.8}$ & 0.314 & $1.1 _{-3.0} ^{+3.6}$ & 0.359\\
   2--4 vs 10--20  & $7.2 _{-7.6} ^{+7.6}$ & 0.172 & $6.0 _{-8.7} ^{+8.6}$ & 0.253\\
   \hline
  \end{tabular}
   \end{center}
   \caption{The estimated soft lags using FR/RSS for CEDCF and ZDCF methods. 
   In this analysis, CEDCFs are calculated with one day intervals and $n_{min}$ is confined to 11 for calculating ZDCFs. 
   The details of these methods are written in the text.}
   \label{tab:cent_all}
\end{table}

\subsection{Analysis of the flux enhancement event}

\subsubsection{Light Curves}
On the lag analysis described above, we found that the strong surge of X-ray flux observed during 55075--55150 
contributed to the soft lag suggested in \S4.1. 
For studying this event more reliably, we remade the light curve by using the maximum-likelihood image fitting method 
as described in Hiroi et al. (2013). 
This method allows us to suppress the back ground fluctuation, and thus make a high quality light curve.
We employed the following procedure with the data selection criteria and the background model described in Hiroi et al. (2013). 
We excluded the event data taken at high latitude regions ($|\rm{latitude}|>40^{\circ}$), where the charged particle flux is high, 
and those detected in near both ends of each proportional counter at a photon incident angle $|\phi| > 38^{\circ}$ 
(for the definition of $\phi$, see Mihara et al. 2011) where the background rejection efficiency is lower. 
While this refined background model is available only in a limited to a portion of time in the {\it MAXI} observation 
in energy range of the 2--4 keV and the 10--20 keV at present, 
this method successfully reduces the flux errors attributable to background fluctuations. 
The remade light curves are not significantly different from the one obtained through the {\it MAXI} on-demand process in this epoch 
except for uncertainties in each bin.

Figure \ref{fig:lc} shows the light curve of Cen A with a bin size of one day 
during MJD=55075--55160 in the three energy bands, 
beneath which we show the hardness ratio (the 4--10 keV divided by the 2--4 keV). 
Calculated fluxes binned in 1 day are represented in black filled circles. 
We can see a flux enhancement peaking around MJD $\sim$ 55090 at a flux $\sim$ 60 mCrab in the 4--10 keV band. 
It is the brightest flux recoded by the {\it MAXI}/GSC so far for this object. 
The relative amplitudes of flux enhancements (a factor of $\sim$ 2) are similar in the three energy bands. 
On the other hand, the temporal variability looks somewhat different among the three. 
The flux decay time in the 2--4 keV seems longer than those in the other energy bands, 
while the timescale of variation in the 10--20 keV seems shorter than those in the other energy bands. 
As for the hardness ratio in figure \ref{fig:lc},  
we can see the decreasing trend toward the peak of the flux enhancement event from MJD = 55075 to MJD = 55100, 
and then it stays almost at a constant. 

\subsubsection{Time lag Analysis}
Using the method of the time lag analysis based on the DCF described in \S 3, 
we investigated the correlations and time lags of the X-ray fluctuations among three energy bands 
during the flare observed at MJD $\sim$ 55090. 
The CEDCFs and the ZDCFs between 
the 2--4 keV and 4--10 keV, the 4--10 keV and 10--20 keV, 
and the 2--4 keV and 10--20 keV are shown in Figure \ref{fig:dcf}.
The CEDCFs calculated with a bin size of 2 days and a distance between centers of bins of 1 day are shown by shadowed areas, 
and the data points and the error bars plotted on the shadow areas represent the ZDCFs. 
One can see in Figure \ref{fig:dcf} that the calculated cross-correlation functions through the two methods give almost the same results. 
The centroids of CEDCFs are located at +7.3 days, +6.7 days, and +11.4 days, 
while those of ZDCFs are found at +3.1 days, +1.4 days, and +7.8 days for the three pairs, respectively. 
These results indicate that the existence of soft lags of $\sim$5--10 days in X-ray flux variation from Cen A. 
Additionally, the shift of the DCF centroid to a positive direction ({\it i.e.,} soft lag) 
are caused by the overall shape of the DCF, 
in other words, it is attributable to the trend on a timescale of $\sim$ 20--30 days (estimated from the HWHM of the DCFs), 
rather than to the small fluctuations on a timescale of $\sim$ 3--5 days. 
The soft lags, therefore,  arise not from the rapid fluctuations especially found in the light curve of the 10--20 keV 
but rather from the more slow variations on a similar timescale as the HWHM of DCFs. 
Correspondingly, even if we use the light curve with the bin size of three day for evaluating DCFs, 
the features described above are reproducible.

The statistical uncertainties of the soft lags are derived through the same procedure as that used in\S 4.1, {\it i.e.} the FR/RSS method, 
by changing the bin width from 2 day to 5 day with a step of 1 day. 
For ZDCF, we set the two parameters as $n_{\rm{min}}=11$ and $\epsilon = 1$ day. 
The evaluated results from 4000 Monte-Calro simulations are summarized in Table \ref{tab:cent}. 
Confidence levels of soft lags are significantly improved compared with the case of the whole light curves, 
and reach 99\% in the 2--4 keV vs. 10--20 keV in this period. 
Again, the FR/RSS method yields a somewhat broadened distribution of  the centroid of DCF compared to the true situation. 
Thus, these results, especially the case of the DCF between the 2--4 keV and 10--20 keV, 
suggest that the day-scale soft lags exists in X-ray emission from Cen A 
at the confidence level of potentially $\ge 99\%$ in flux variations on a timescale longer than $\sim$ 20 days. 

\begin{table}
 \label{table:data_type}
 \begin{center}
  \begin{tabular}{lcccc}
   \hline
   Energy band (keV) & $\tau ^{\rm{CEDCF}}_{\rm{cent}}$ (days) & $P^{\rm{CEDCF}}_{<0 \ (\rm{ days})}$
   & $\tau ^{\rm{ZDCF}}_{cent}$ (day) & $P^{\rm{ZDCF}}_{<0 \ (\rm{ days})}$ \\
   \hline \hline
   2--4 vs 4--10  & $6.7 _{-4.4} ^{+4.7}$ & 0.072 & $4.9 _{-3.5} ^{+3.1}$ & 0.089\\
   4--10 vs 10--20  & $6.3 _{-4.8} ^{+4.9}$ & 0.104 & $3.4 _{-4.0} ^{+4.2}$ & 0.202\\
   2--4 vs 10--20  & $12.2 _{-4.9} ^{+5.2}$ & 0.008 & $9.0 _{-3.9} ^{+4.8}$ & 0.011\\
   \hline
  \end{tabular}
   \end{center}
   \caption{The estimated soft lags during MJD = 55075--55160 using FR/RSS for CEDCF and ZDCF methods. 
   In this analysis, CEDCFs are calculated with one day intervals and $n_{min}$ is confined to 11 for calculating ZDCFs. 
   The details of these methods are written in the text.}
   \label{tab:cent}
\end{table}

\section{Discussion}
We have found a soft lag of $\sim$ 5--10 days in Cen A in the X-ray flux variations on a timescales longer than $\sim$ 20 days 
between the 2--4 keV and 10--20 keV bands with a confidence level of 99\% in the flux enhancement event at MJD $\sim$ 55090. 
It could be argued that the high significance of the soft lag in the first flare episode should be discounted 
because the significance in the later episode are not so high. 
We note, however, that the first episode had the highest peak flux in the 4--10 keV (and also 2--20 keV), 
and that the sensitivity of the {\it MAXI}/GSC was best at that epoch. 
While it is difficult to demonstrate that a soft lag is always present in every flux enhancement events, 
the boxcar CEDCF for the later episode in Figure \ref{fig:box_dcf} shows 
a similar feature to that of the first event, and suggestive of the existence of a soft lag.
In the following sections, we discuss if the observed features described above are accountable by the 
X-ray emission or reprocess scenarios for AGNs; namely, 
the synchrotron cooling of electrons in the one-zone synchrotron self-Compton process in the jet, 
cooling of electrons by the Compton up-scattering of the disk photons in the corona, 
and X-ray reverberation on the accretion disk or a warm absorber. 

\subsection{One-zone synchrotron self-Compton}
Soft lags in AGNs are sometimes explained by the radiative cooling of relativistic electrons 
through the one-zone synchrotron self-Compton (SSC) process ({\it e.g.} Kataoka et al. 2000).
In this model, the power produced by a single electron in the synchrotron radiation and the inverse Compton scattering 
of a single electron is given by (Rybicki \& Lightman 1979)
\begin{equation}
P = \frac{\mathrm{d}}{\mathrm{d} t} [(\gamma-1) m_{\rm{e}} c^2] = \frac{4}{3} c \sigma_{\rm{T}} (\gamma^2 -1) (U_{\rm{B}} + U_{\rm{ph}}), 
\end{equation}
where $m_{\rm{e}}$ is the mass of an electron, $\sigma_{\rm{T}}$ is the Thomson cross section, 
and $U_{\rm{B}}$ and $U_{\rm{ph}}$ are the energy densities of magnetic field ($B^2/8\pi$) and the seed photon field, respectively. 
The cooling timescale, which sets the characteristic timescale for the kinetic energy loss of electrons by 
the synchrotron radiation and the inverse Compton scattering, can be expressed as follows:
\begin{equation}
t_{\rm{cool}} = \frac{(\gamma-1) m_{\rm{e}} c^2}{4 c \sigma_{\rm{T}} (\gamma^2 -1) (U_{\rm{B}} + U_{\rm{ph}}) / 3} 
= \frac{3 m_{\rm{e}} c}{4 \sigma_{\rm{T}} (\gamma+1) (U_{\rm{B}} + U_{\rm{ph}})}.
\end{equation}
In the jet comoving frame, 
the time lag ($t_{\rm{lag}}$) of emission at $\gamma_2$ with respect to emission at $\gamma_1$ is 
the time it takes for electrons to lose energy equal to $(\gamma_1-\gamma_2)m_{\rm e}c^2$. 
Thus, $t_{\rm{lag}}$ in the jet comoving frame is given by
\begin{equation}
t_{\rm{lag}} 
= \frac{3 m_{\rm{e}} c}{4 \sigma_{\rm{T}} (U_{\rm{B}} + U_{\rm{ph}}) } \left( \frac{1}{\gamma_2+1} - \frac{1}{\gamma_1+1} \right). 
\end{equation} 
In this paper, we consider the upper limit of the cooling time of an electron:
\begin{equation}
t_{\rm{lag}} < \frac{3 m_{\rm{e}} c}{4 \sigma_{\rm{T}} U_{\rm{B}} } \frac{1}{\gamma_{\rm{min}}+1},
\end{equation} 
where $\gamma_{\rm{min}}$ is the minimum Lorentz factor of electrons in the emission region.
Abdo et al. (2010) attempted to explain the multi-wavelength SED of Cen A with the one zone SSC model 
and yielded the magnetic field of the emitting region $B = 6.2$ G, a minimum electron Lorentz factor $\gamma_{\rm{min}} = 300$, 
and the doppler factor $\delta = 1$. 
Assigning these values to Eq.(12) and considering that the cooling time in the observer's frame is not different in the comoving frame 
under the situation of $\delta = 1$, 
we obtain $< 1.1$ day as a conservative upper limit of radiative cooling time of the electrons in the observer's frame. 
The upper limit seems 
only marginally consistent with the observed day-scale soft lags 
within flux variations on a timescale $\gtrsim 20$ days, 
if the observed flux enhancement is caused by a single energy injection. 
One way to produce a variation timescale longer than an intrinsic electron cooling time is 
the emission region with a light-crossing time longer than the $t_{{\rm cool}}$. 
In this case, however, the shape of light curves expected be close to symmetric 
as suggested in Kataoka et al. (2000), in contradiction to the X-ray light curves shown in figure \ref{fig:lc}. 
Additionally, an emission region size of $\sim$20 light days 
corresponding to $\sim$850 Schwarzschild radius ($R_{\rm{s}}$) for the black hole mass 
of  $M_{\rm{BH}} = 2.0 \times 10^8 M_{\odot}$ (Marconi et al. 2001) is required to construct such a condition. 
It seems to be too large for the emitting region size $\sim 3\times 10^{15}$ cm (corresponding to $\sim 50 R_{\rm{s}}$) 
estimated by fitting the broad-band SED (Abdo et al. 2010). 
If the enhancement of flux intensity with soft lags was triggered by a single energy injection, therefore, 
a weaker $B$ field or/and a small $\gamma_{\rm{min}}$ are necessary at the period. 

Alternatively, multiple injections may be able to produce asymmetry in a CEDCF 
and a soft lag longer than the $t_{{\rm cool}}$ at the same time 
if the time resolution and photon statistics in the present data is not sufficient to resolve individual flares. 
For example, hard-to-soft evolution in hardness ratio is often observed in gamma-ray bursts (GRBs) with multiple pulses, 
both within a single pulse and over series of pulses in a burst ({\it e.g.} Fishman \& Meegan 1995, Fenimore et al. 1995). 
If multiple energy injections in a homogeneous region ({\it e.g.} series of internal shocks caused by repeated ejection of matter), 
or coincidentally discrete single energy injections in several regions, 
occurs in Cen A and undergoes hard-to-soft evolution, 
a soft lag could be observed. 
In fact, a decreasing trend in the hardness ratio ({\it i.e} the softening trend) is suggested 
around the peak of the flux in Figure \ref{fig:lc}. 
SSC process with a multiple injection scenario is, therefore, not rejected by our results. 

\subsection{Up-scattered black body radiation}
The X-ray emission in a Seyfert galaxy is commonly thought to be produced by inverse Compton scattering 
in a hot corona surrounding the inner parts of the accretion disk. 
Here we discuss the reasonable situation that can yield the day-scale soft lag in 
variations on a timescale $\sim$ 20 days on the basis of this assumption. 

In the Comptonization process, the variation in flux may be caused either 
by the seed photon or by the Comptonizing plasma. 
The former case, naturally leads to hard lag 
associated with the propagation of mass accretion fluctuations. 
A soft lag, on the contrary, may be caused by the cooling of Comptonizing plasma. 
However, 
the hot corona implied by 
the main X-ray power-law continuum which extends to hundreds of keV, 
seen in {\it INTEGRAL} data of Cen A (Beckmann et al. 2011, Burke et al. 2014) , 
is not relevant to produce variations with soft lag in the soft X-ray region (2--20 keV). 
To produce the observed soft lag, the plasma temperature (or a spectral break energy) 
needs to be around $\sim$ 10 keV. 
However, we think it difficult to expect the temperature of the hot corona responsible for the X-ray continuum 
to decrease from hundreds of keV to $\lesssim$ 10 keV considering the stability 
in the X-ray band as observed by {\it RXTE} (Rothschild et al. 2011). 
Such a low-temperature corona, alternatively, is possibly present in addition to the hot corona mentioned above, 
and its cooling may cause the observed soft lag. 
This idea has been applied to explain the 
Comptonized component with a low break energy 
found in a Seyfert galaxy Mrk 509 (Noda et al. 2011). 
They found that a soft excess component detectable at $\lesssim$ 3 keV 
varies independently of the main X-ray continuum, 
and argued that this soft component with $kT_{e} = 0.49$ keV 
is a evidence for ``multi-zone Comptonization'' 
similar to the case of the black-hole binary Cygnus X-1. 
In Mrk 509, the soft excess component was stable during a few days 
but varied on timescales of a few weeks. 
If Cen A has similar multi-zone Comptonization component 
and the soft component has the plasma temperature of several keV, 
observed soft lag may be explained by cooling of the Comptonizing plasma. 
The nuclear emission of Cen A is not detectable in optical or UV spectrum, 
due to a high column density ($N_{\rm H} \sim 1.6 \times 10^{23}$ atms/cm$^2$; Rivers et al. 2011). 
Therefore, we do not have good knowledge of the seed photon 
from the inner accretion disk, which influence the cooling time. 
The cooling time of the Comptoniazation plasma in the non-relativistic limit, however, 
is suggested to correspond to the power spectrum density (PSD) break in the X-ray band 
of Seyfert galaxies (Ishibashi and Courvoisier, 2012). 
For Cen A, Rothschild et al. (2011) measured the PSD break in 1.5--12 keV band 
at $T_{\rm{b}}= 18.3 ^{+18.3}_{-6.7}$ days. 
It is consistent with the timescale of the variation in our results.
These argument supports our interpretation that the observed soft lag 
is due to a soft component of multi-zone Comptonization. 

\subsection{Reverberation on the Accretion Disk, Warm Absorbers, the BLR, or the Torus}
Previous observations have revealed that X-ray spectra of AGNs often show a soft X-ray excess below $\sim$ 1 keV, 
superposing above a main power law component. 
This phenomenon is usually interpreted as the reflection paradigm 
where a corona (power-law component) irradiates 
the warm absorber lying around $100 R_{\rm{s}}$ (corresponding to $\sim 2.3$ light days for Cen A) 
from the black hole ({\it e.g.} Miller et al. 2008 ), 
or the inner region of the accretion disk within $20 R_{\rm{s}}$ (corresponding to $\lesssim 0.46$ light days for Cen A) where relativistic blurring is significant ({\it e.g.} Cackett et al. 2013). 
In the both hypotheses, the soft excess is naturally explained as the contribution of Fe L lines in these origin. 

Recently, De Marco et al. (2013) reported lags of soft excess with respect to the X-ray continuum in over a dozen AGNs, 
and explained that the delayed soft excess emission originates in the innermost regions of the accretion disc. 
Our data, however, starts at 2 keV and do not contain the energy region containing the Fe L lines. 
This model, therefore, predict a hard lag in contradiction to our results on the contrary. 
With reflection, the variation of the 2--4 keV continuum flux that represents the direct component 
lead the variation in the 4--10 keV range containing the Fe K line and the 10--20 keV band containing the Compton hump. 
In fact, Zoghbi et al. (2014) reported a time delay between energies $> 10$ keV and the continuum in MCG--5-23-16. 
They claimed that the these lags are most likely due to reverberation in the reflection Compton hump 
in a manner similar to the response of the relativistic Fe K line. 
Besides, a reflection component from the broad line region (BLR) or the torus 
located at a range of a mili-parsec to a parsec 
(a few light days to a few light years; {\it e.g.} Peterson 1993, Krolik and Begelman 1988) from the center 
would be detected in the X-ray spectrum as a hard lagging component. 
Therefore, the reverberation process on the accretion disk, warm absorbers, the BLR or the torus 
cannot explain the soft lags observed in Cen A. 

\section{Summary}
Employing the DCF methods, we found evidence for soft lags on a timescale of days 
in the flux enhancement event from MJD 55075 to MJD 55150, 
between all energy bands' X-ray variation in 2--4 keV, 4--10 keV, and 10--20 keV of Centaurus A 
in the X-ray light curve observed by the {\it MAXI}/GSC, 
and quantify the significance of these results by performing Monte-Carlo simulations known as the FR/RSS method. 
The confidence of existence of the day-scale soft lag is higher than 99\%. 
Also, we find that the soft lag is mainly originated by flux variations on a timescale of $\gtrsim 20$ days. 

If the observed X-rays are emitted by SSC in a single zone in the jet 
by a single injection
with the parameters 
: $B$ = 6.2 G, $\gamma_{\rm{min}}=300$, and $\delta$ = 1 given by Abdo et al. (2010), 
the day-scale soft lags is difficult to explain because of too short cooling time of highly energetic electrons. 
For interpreting the evaluated soft lags by the simple one zone SSC model, 
a multiple injection scenarios with a hard-to-soft evolution, or 
a weaker $B$ field or/and a smaller Lorentz factor of electrons are required.
Alternatively, assuming that the variable X-rays originate in the disk photons which are up-scattered by electrons in a corona 
with a temperature of around 10 keV 
surrounding the inner part of the accretion disk, 
we show the possibility that the observed lags can be explained by the cooling  
of the Comptonizing corona. 
The plausibility of this scenario is supported by the consistency in the decay times and the break timescale of the power spectrum density. 
With the reverberation scenario, a hard lag is naturally  expected, and therefore the observed soft lags cannot be explained. 



\newpage

\begin{figure} 
 \begin{center}
  \includegraphics[width=15cm]{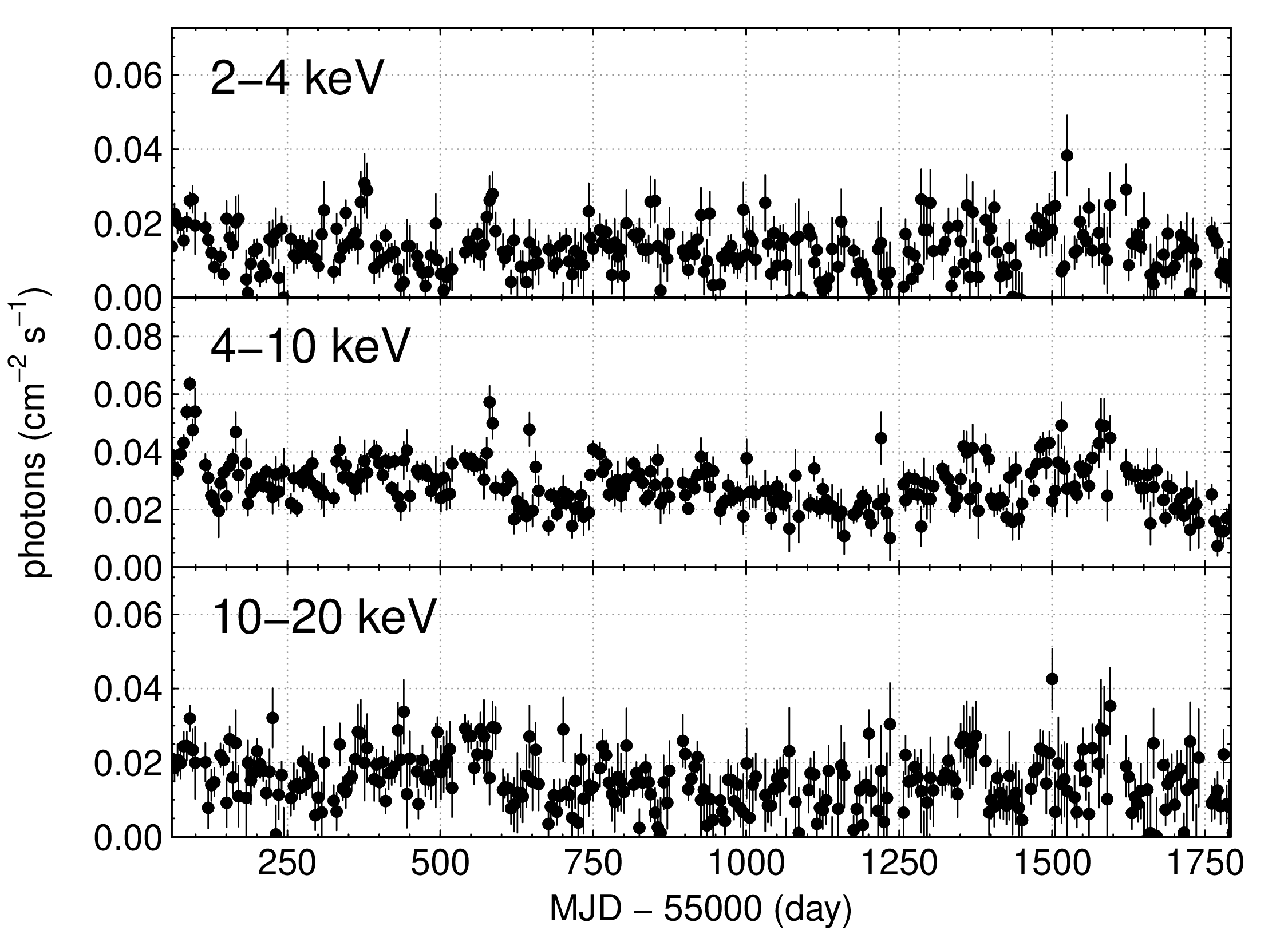} 
 \end{center}
\caption{The five days binned long-term X-ray light curves of Centaurus A in the 2--4 keV (top panel), 4--10 keV (middle panel), and 10--20 keV (bottom panel) 
obtained by the {\it MAXI}/GSC. }\label{fig:lc_all}
\end{figure}

\begin{figure} 
 \begin{center}
  \includegraphics[width=15cm]{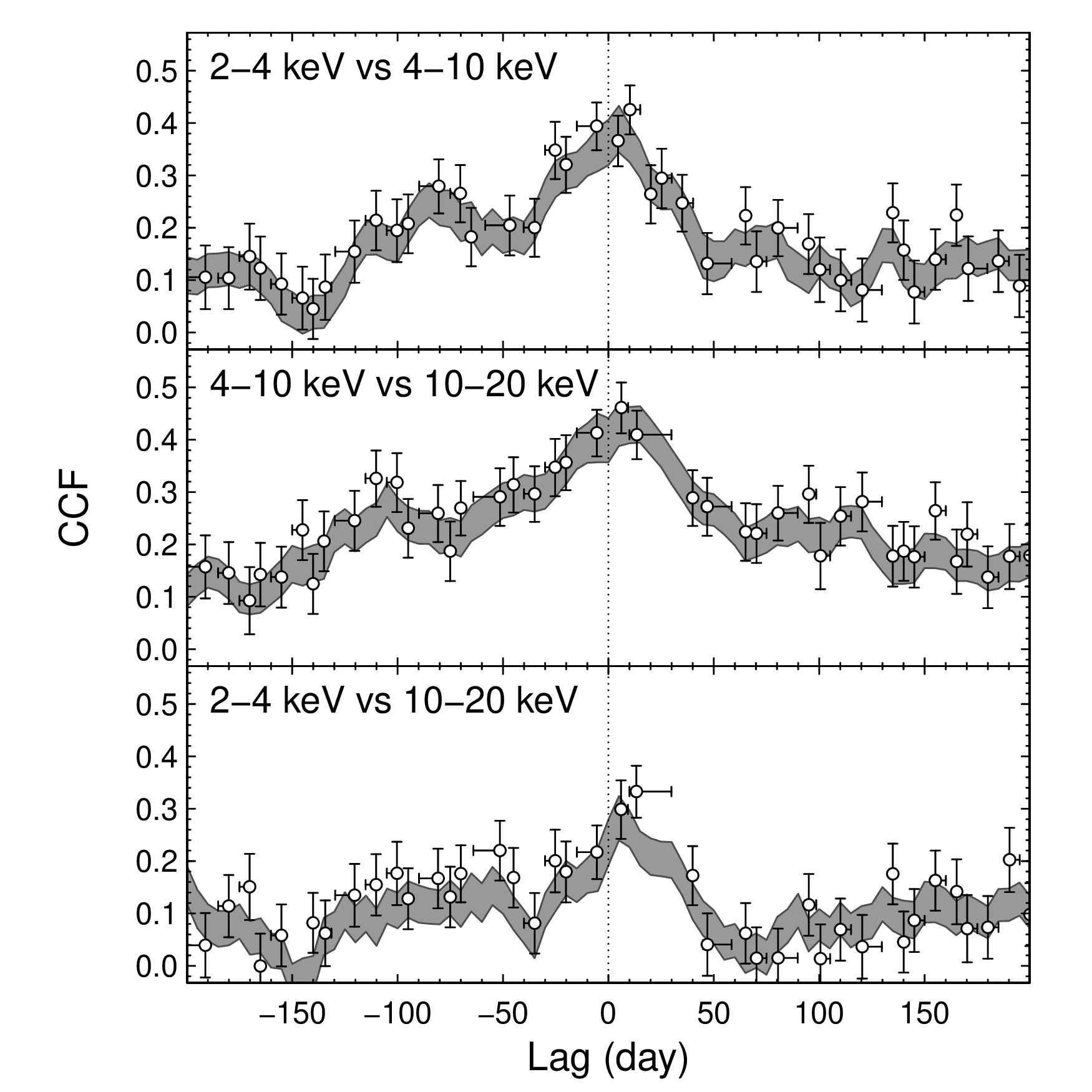} 
 \end{center}
\caption{CEDCFs and ZDCFs calculated between the 2--4 keV and 4--10 keV, 4--10 keV and 10--20 keV, and 2--4 keV and 10--20 keV 
  for the long-term light curves. 
  CEDCFs are shown as gray shadowed area, and ZDCFs are plotted by black circles with error bars. 
  Dashed vertical line stands for $\tau_{\rm{lag}} = 0$. 
  For the definition of the horizontal error bars, see \S 3.3.}\label{fig:dcf_all}
\end{figure}

\begin{figure} 
 \begin{center}
  \includegraphics[width=15cm]{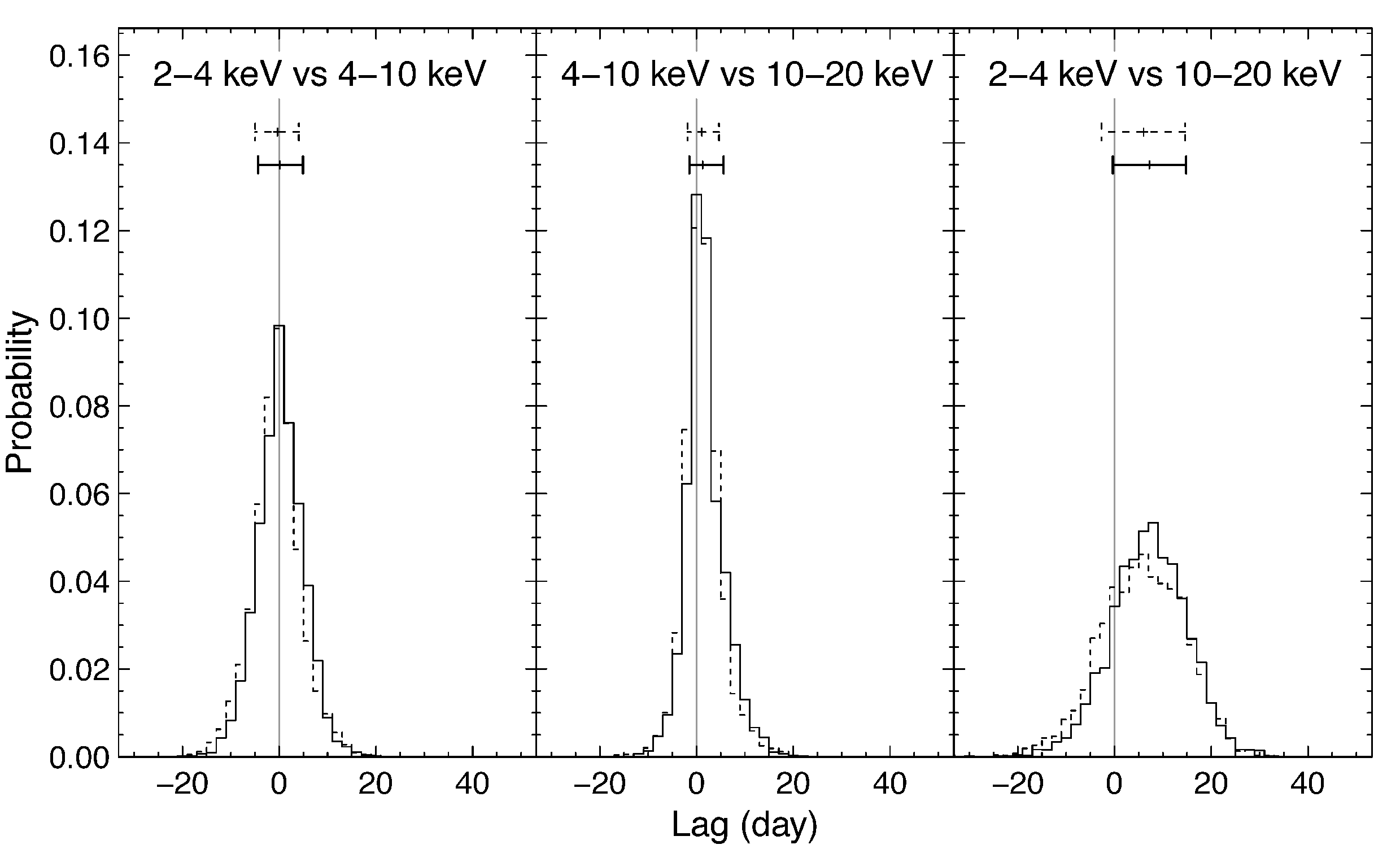} 
 \end{center}
\caption{Cross correlation centroid distributions (CCCDs) of CEDCF (solid line) and ZDCF (dashed line) for the long-term light curves 
  obtained by the FR/RSS method. 
  Derived soft lags and corresponding 1$\sigma$ ranges are also shown above the histograms. 
  Gray vertical lines are drawn at $\tau_{\rm{lag}} = 0$. }\label{fig:hist_all}
\end{figure}

\begin{figure} 
 \begin{center}
 \hspace*{-2.5em}
  \includegraphics[width=15cm]{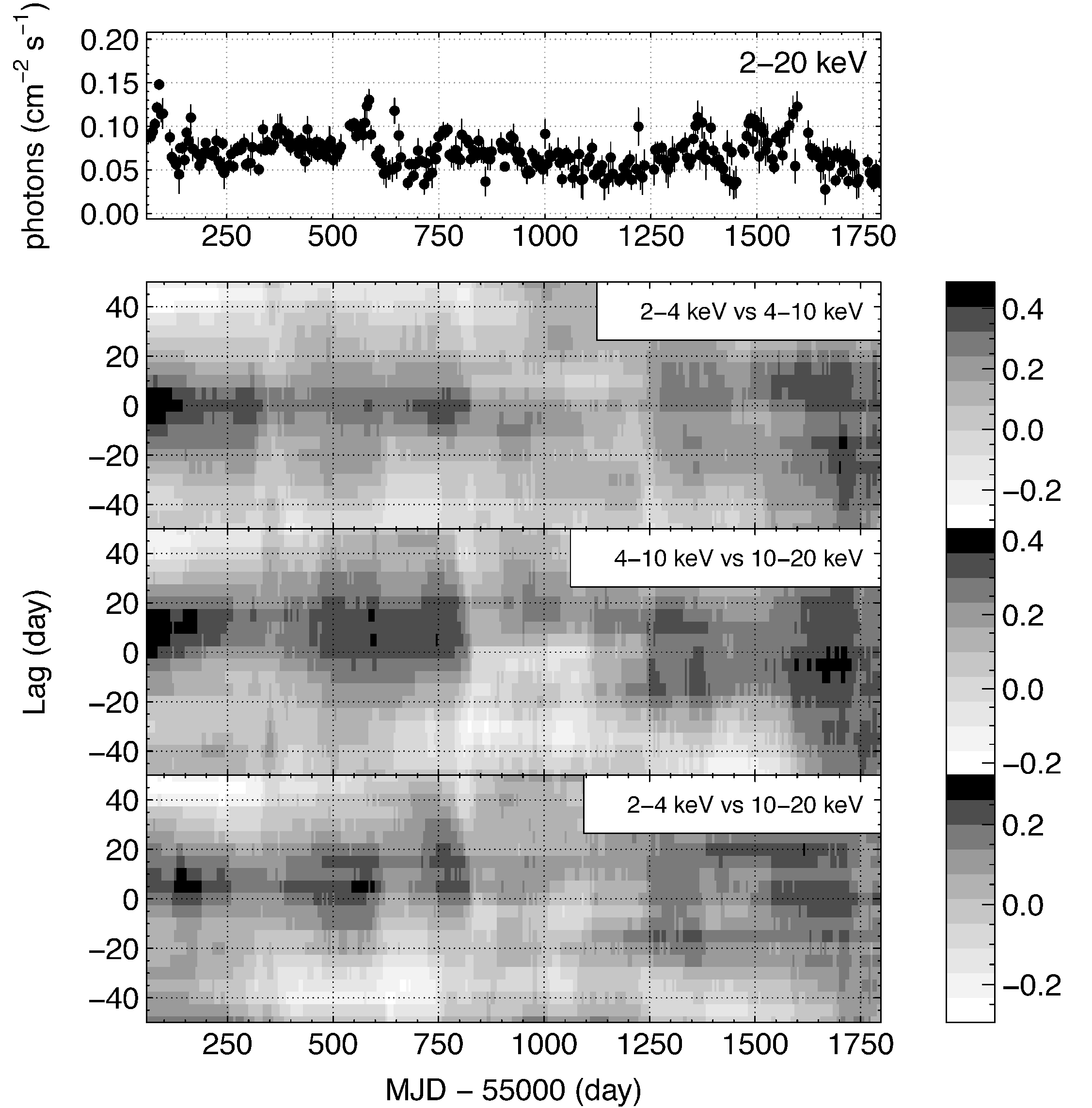} 
 \end{center}
\caption{Running CEDEFs for full length X-ray light curves between 2-4 keV and 4-10 keV, 4-10 keV and 10-20 keV, and 2-4 keV and 10-20 keV. The gray scale shows a correlation coefficient between different energy bands' light curves separated by 200days. The top panel is the 5 days binned light curve for reference.}\label{fig:box_dcf}
\end{figure}

\begin{figure} 
 \begin{center}
  \includegraphics[width=15cm]{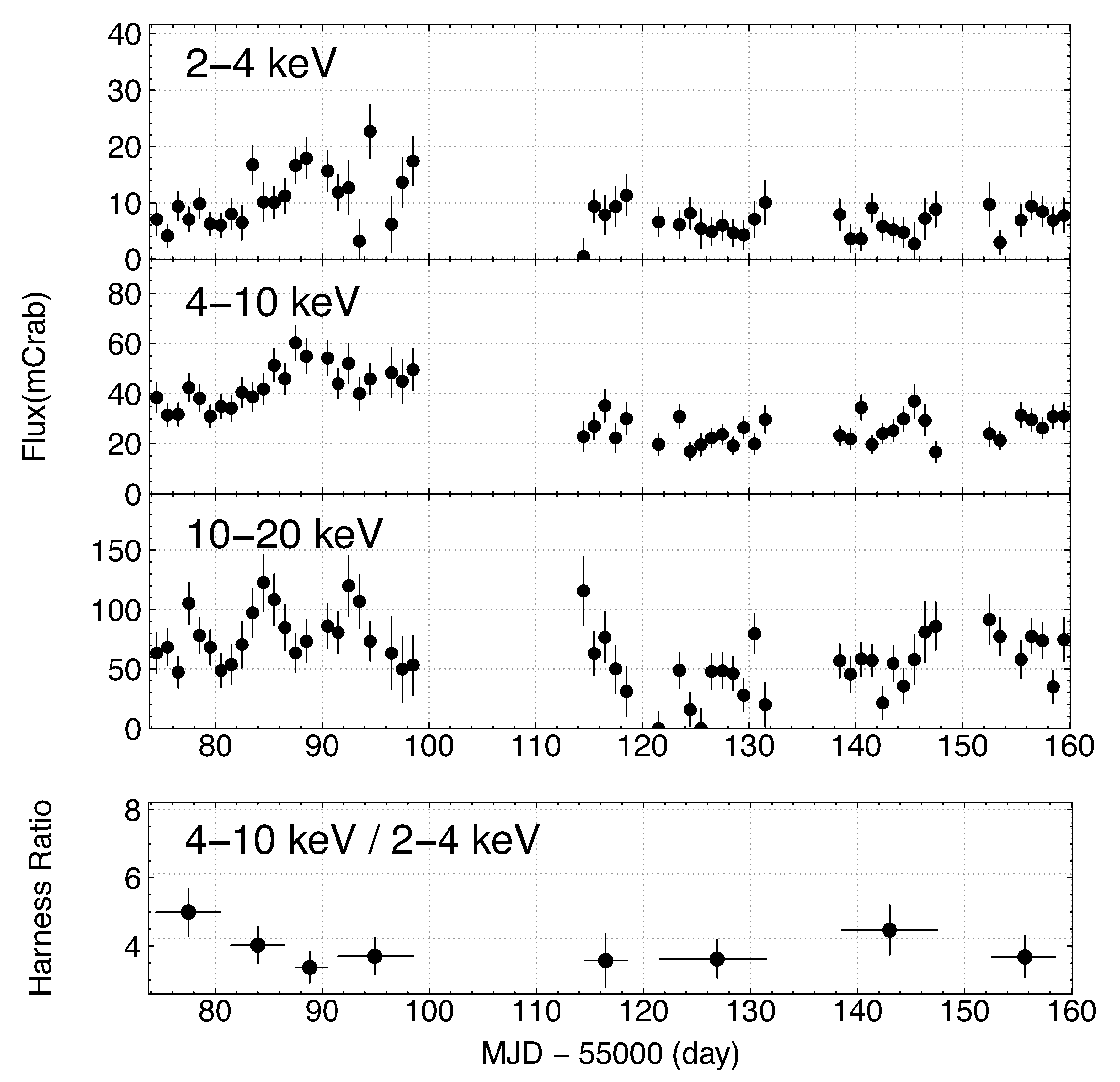} 
 \end{center}
  \caption{X-ray light curves of Centaurus A obtained by the MAXI/GSC. 
   The flux data are plotted as the black points with error bars. 
   The error bars present 1 $\sigma$ errors derived from $C$-statistics (see Hiroi et al. 2011 for details). 
   The bottom panel is the ratio of the 4--10 keV to the 2--4 keV flux.}\label{fig:lc}
\end{figure}

\begin{figure} 
 \begin{center}
  \includegraphics[width=15cm]{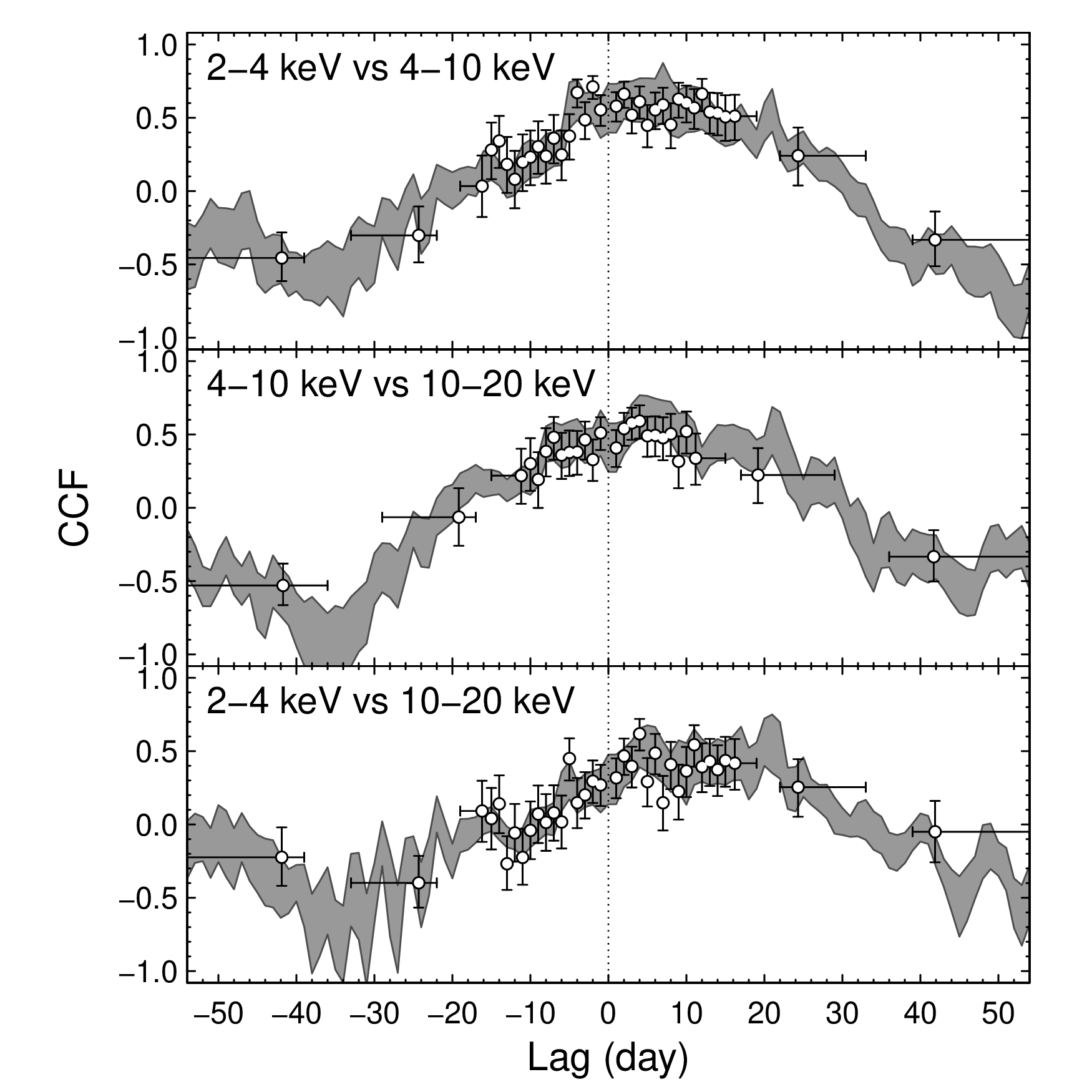} 
 \end{center}
  \caption{CEDCFs and ZDCFs calculated between the 2--4 keV and 4--10 keV, 4--10 keV and 10--20 keV, and 2--4 keV and 10--20 keV 
  for the flux enhanced episode. 
  CEDCFs are shown as gray shadowed area, and ZDCFs are plotted by black circles with error bars. 
  Dashed vertical line stands for $\tau_{\rm{lag}} = 0$. 
  For the definition of the horizontal error bars, see \S 3.3.}\label{fig:dcf}
\end{figure}

\begin{figure} 
 \begin{center}
  \includegraphics[width=15cm]{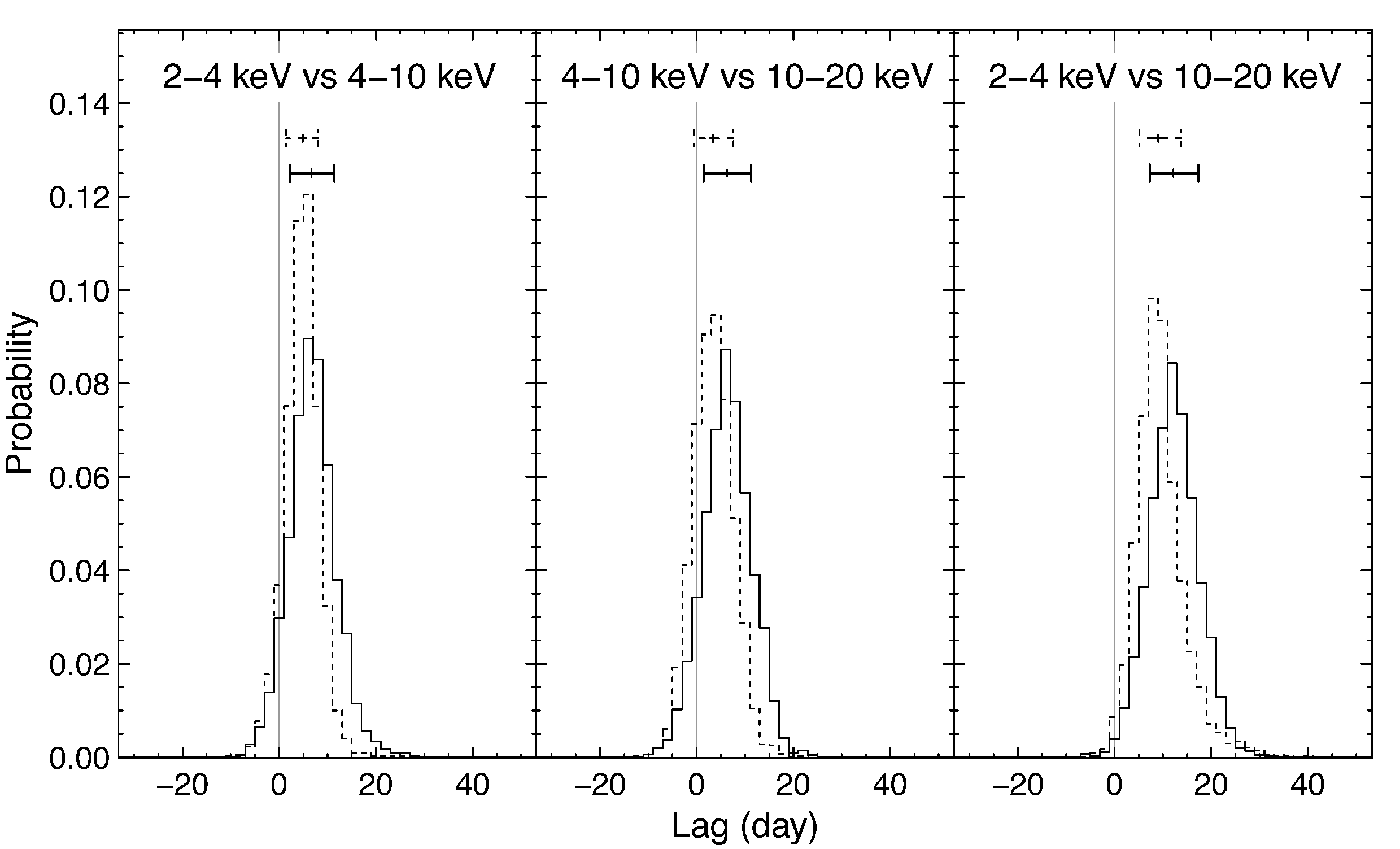} 
 \end{center}
  \caption{Cross correlation centroid distributions (CCCDs) of CEDCF (solid line) and ZDCF (dashed line) for the flux enhancement event 
  obtained by the FR/RSS method. 
  Derived soft lags and corresponding 1$\sigma$ ranges are also shown above the histograms. 
  Gray vertical lines are drawn at $\tau_{\rm{lag}} = 0$. }\label{fig:hist}
\end{figure}


\end{document}